\documentstyle[11pt,newpasp,twoside,epsf]{article}
\newcommand{\HI}{\protect\normalsize H\thinspace\protect\footnotesize
I\protect\normalsize}

\def\edcomment#1{\iffalse\marginpar{\raggedright\sl#1\/}\else\relax\fi}
\reversemarginpar

\begin{document}
\title{An H\,{\small \bf I} Survey of the Great Attractor Region} 

\author{Lister Staveley-Smith,$^1$ Sebastian Juraszek$^{1,2}$
Patricia A. Henning$^{3}$, \\ B\"arbel S. Koribalski$^1$, 
Ren\'ee C. Kraan-Korteweg$^4$}
 
\affil{$^1$Australia Telescope National Facility, CSIRO, PO Box 76,
  Epping, NSW 1710, Australia}

\affil{$^2$School of Physics, University of Sydney, NSW 2006, Australia}

\affil{$^3$Institute for Astrophysics, University of New Mexico, 800
Yale Boulevard, NE, Albuquerque, NM 87131, USA}

\affil{$^4$Depto.\ de Astronom\' \i a, Universidad de Guanajuato, 
Apdo.~Postal 144, Guanajuato GTO 36000, Mexico}

\begin{abstract}
  A blind \HI~survey using the Parkes telescope at $|b|<5\deg$ $300\deg
  < \ell < 332\deg$ has so far revealed 305 galaxies, most of which
  were previously unknown. These galaxies are used to map out the
  distribution of filaments and voids out to $10^4$ km s$^{-1}$. A
  preliminary measurement of the galaxy overdensity suggests only a
  moderate overdensity is present, and that the excess mass (above the
  background density) is $\sim2\times10^{15} \Omega_{\circ}$
  M$_{\odot}$. This is below the mass predicted in POTENT
  reconstructions of the local velocity field, and implies that the
  `Great Attractor' (GA) is not as massive as these reconstructions
  indicate, or does not lie hidden in the region investigated.
\end{abstract}

\section{Introduction}

As described elsewhere in these proceedings, the Zone of Avoidance
(ZOA) provides a formidable obstacle to the study of galaxies and to
the completion of our picture of large-scale structure in the local
Universe. The southern ZOA obscures dynamically important structures
in the Puppis, Vela, Norma and Ophiuchus regions (Lahav 1994,
Kraan-Korteweg \& Woudt 1994, Kraan-Korteweg et al.~1996, Wakamatsu et
al.\ 1994). Density fields inferred from complete redshift surveys
such as PSCz (Saunders et al., these proceedings) and peculiar
velocity reconstructions (Kolatt, Dekel, \& Lahav 1995) indicate that
the most massive structure, usually dubbed the `Great Attractor' (GA),
probably lies near the Galactic equator close to $\ell=320\deg$.

Surveys of galaxies down to moderately low latitudes are possible with
some difficulty at optical and infrared wavelengths. However, quantitative
studies which involve measurements of luminosity or size are very hard
to make with any precision because of the obscuration problem. The blind
21-cm approach as pioneered by Kerr \& Henning (1987) offers a neat
solution to the problem, in that flux densities are unaffected by dust.
Moreover, with the \HI~Parkes All-Sky Survey (\HI\/PASS) now complete, 
measurements of galaxy overdensity in previously obscured regions can be
compared with very large control samples.

This paper describes some preliminary results from a deep survey with
the multibeam receiver on the Parkes telescope. Over 300 galaxies were
detected in this survey, which is an extension of the survey begun by
Juraszek et al. (2000). The survey is dense enough to recognise several
coherent structures and to begin to measure actual overdensities in order
to quantify excess mass in this region.

\section{The Survey}

\begin{figure}[htb]

    \plotfiddle{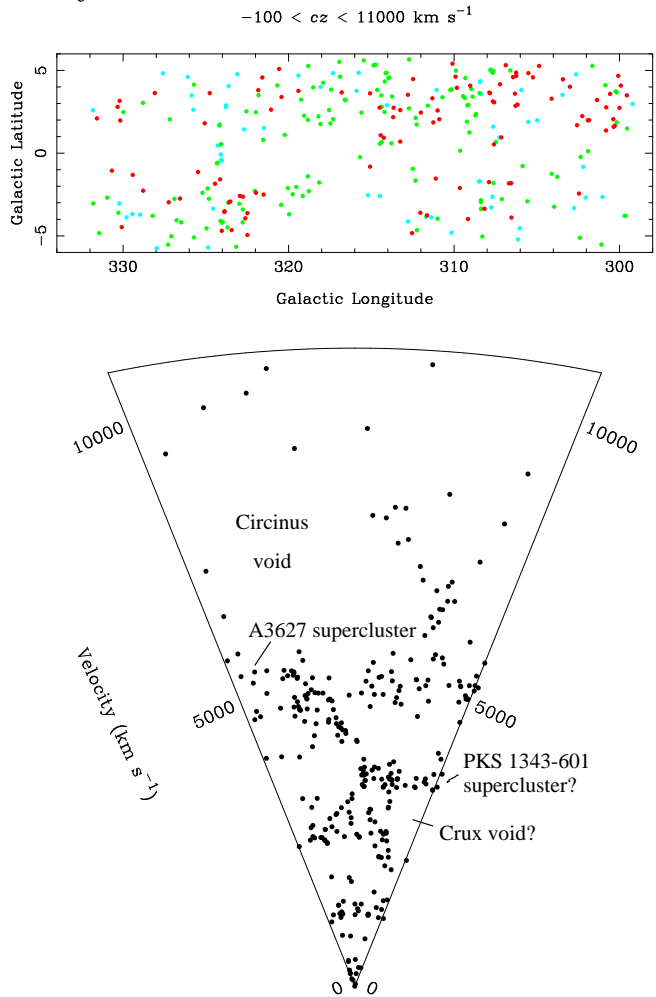}{12cm}{0.0}{100.0}{100.0}{-300.0}{-240.0}
    \caption{{\it Top:} angular distribution of galaxies $|b|<5\deg$ and
      $300\deg < \ell < 332\deg$. {\it Bottom:} wedge diagram showing
      velocity (Local Group frame) as a function of longitude. Several
      coherent structures are identified, following Fairall (1998).}

\end{figure}

\begin{figure}[htb]
    \plotfiddle{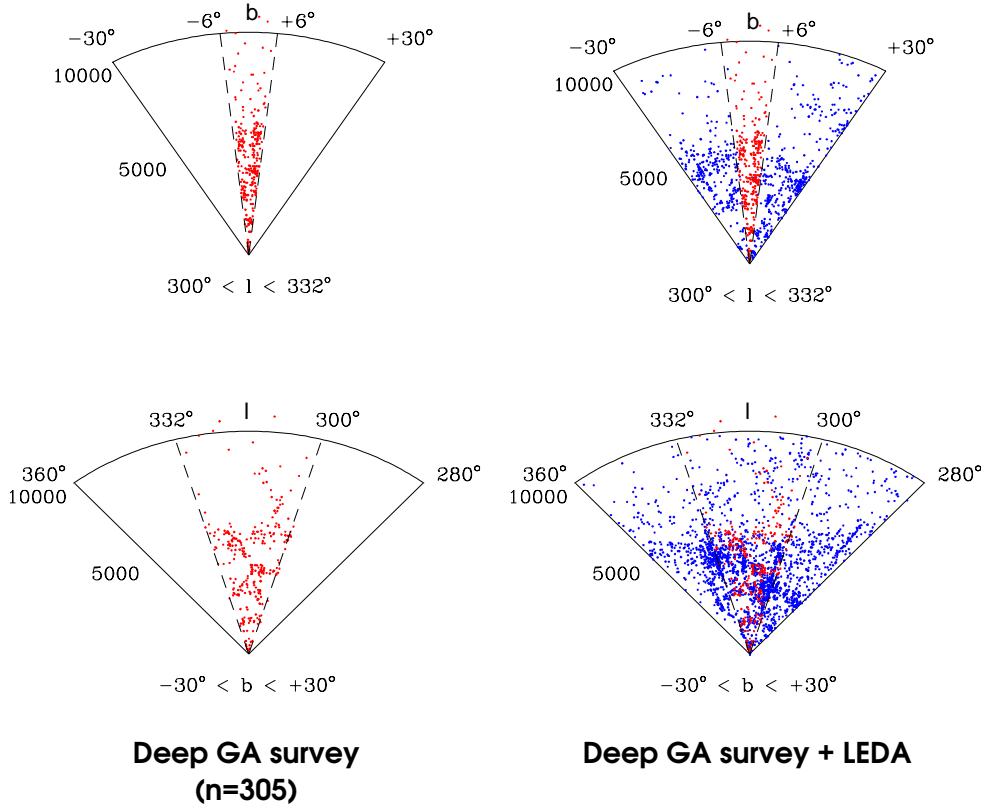}{10cm}{270.0}{60.0}{60.0}{-250.0}{340.0}
    \caption{{\it Left:} distribution of the newly detected galaxies in 
Galactic latitude and Galactic longitude. {\it Right:} 
including LEDA galaxies with known redshifts.}

\end{figure}

The \HI~survey presented here is part of the ongoing Parkes ZOA survey
(Staveley-Smith et al. 1998, Juraszek et al. 2000, Henning et al.
2000).  The aim is to cover the southern ZOA to deeper levels than
afforded by \HI\/PASS alone. The data
presented here comprises the full-sensitivity data for the region
$300\deg < \ell < 332\deg$ and $|b|<5\deg$, or nearly a fifth of the
southern ZOA.

The total integration time is 230 hrs, with the final sensitivity
being 5 mJy beam$^{-1}$. The angular and velocity resolutions are
15\farcm5 and 18 km s$^{-1}$, respectively. The velocity coverage is
$-1200$ to 12700 km s$^{-1}$.

A total of 305 galaxies were found by visual inspection of the data
cubes from two independent searches (SJ and RKK). The cubes have some
residual baseline ripple near continuum sources, which it is planned
to correct for at a later stage. The limiting flux density for
inclusion in the catalogue was dependent on the velocity width of the
\HI~profile approximately according to $W_{50}^{-0.7} \int S dV >
0.05$. In other words, a galaxy with a profile width $W_{50} = 150$ km
s$^{-1}$ will be included if its flux integral exceeds 1.7 Jy km
s$^{-1}$.

\section{Galaxy Distribution}

The distribution of the 305 galaxies on the sky and in redshift is shown in 
Fig.~1. There is clear structure in the galaxy distribution,
with several coherent features visible. There appear to be at least
two voids both of which (Circinus and Crux) are identified by Fairall (1998)
from existing galaxy redshift surveys above and beneath the Plane. At
$\ell \approx 330\deg$, $cz = 5000$ km s$^{-1}$ there is a large overdensity
of galaxies. This redshift coincides closely with the ACO 3627 cluster, and its
spiral-rich subcluster (Woudt 1998), suggesting that the structure is part
of an extended supercluster or cloud of galaxies around the main cluster.
This whole structure appears to be the low-redshift boundary of the
Circinus void and to extend at least $50 h^{-1}$ Mpc to $\ell \approx 300\deg$.

Extending between $\ell = 300\deg$ and $\ell = 314\deg$ at $cz = 3400$
km s$^{-1}$ (Local Group frame), and forming the high redshift side of
the Crux void is a coherent structure which may be associated with PKS
B1343-601, a radio galaxy with a Local Group redshift of 3641 km
s$^{-1}$ (West \& Tarenghi 1989). The existence of a coherent structure
at this redshift would tend to confirm the existence of a cluster around 
PKS B1343-601 as suggested by Kraan-Korteweg \& Woudt (1999).  
We see no direct evidence for the actual cluster, though it may be 
spiral-poor.

\begin{figure}[htb]
    \plotone{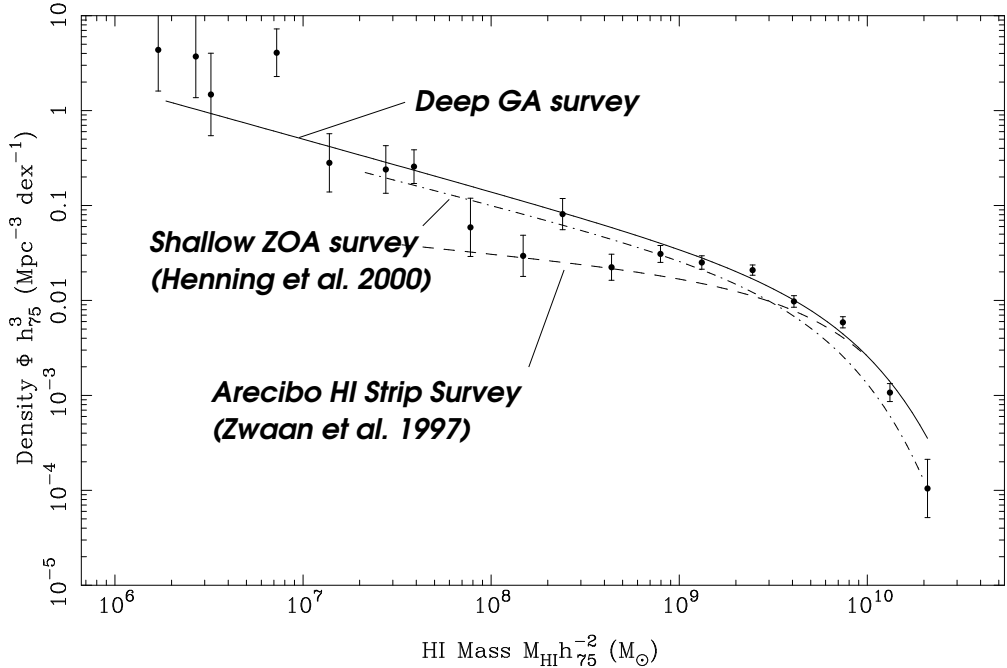}
    \caption{The \HI~mass function of the galaxies in the GA region formed
using the $\Sigma V_{\rm max}^{-1}$ method. A Schechter function
(solid line) is fit to the data. An overdensity of galaxies 
relative to the Zwaan et al. (1997) (dashed line) and Henning et al. (2000) 
(dot-dashed line) \HI~mass functions is apparent.}
\end{figure}

Figure~2 shows the distribution of the same galaxies in
latitude-velocity and longitude-velocity wedges, similar to those
presented by Juraszek et al. (2000). The plots on the right include
galaxies in the Lyon/Meudon Extragalactic Database and together with
the multibeam galaxies give the most complete census of nearby galaxies in the
range $\ell = 300\deg$ to $332\deg$, $|b|<30\deg$.

\section{Mass Overdensity}

The ability to measure \HI~fluxes and masses without uncertain extinction
corrections means that, assuming neutral gas traces total mass on a
large scale, we can measure densities and masses. A straightforward way
of doing this is to measure and integrate the \HI~mass function using
the selection function noted above. 

In Fig.~3, we calculate the volume density for galaxies in a given \HI~
mass interval (0.25 dex bins) using the $\Sigma V_{\rm max}^{-1}$
method.  The resultant mass function is fitted with a Schechter
function, and compared with the results from the shallow ZOA survey of
Henning et al.  (2000) and the Arecibo \HI~Strip Survey (A\HI\/SS) of
Zwaan et al. (1997). The present `deep GA'
survey shows a mass excess compared with both other surveys. This is
particularly true at low \HI~masses where there is an almost order of
magnitude disagreement between the present results and Zwaan et al.
(1997)\footnote{In the low \HI-mass regime, which does not
  dominate the total \HI~mass, the $\Sigma V_{\rm max}^{-1}$ method is
  sensitive to local overdensities.}. Integrating the data points
in Fig.~3 yields a total \HI~density of $8.0\times10^7$ M$_{\sun}$ Mpc$^{-3}$,
compared with $5.9\times10^7$ M$_{\sun}$ Mpc$^{-3}$ for the shallow ZOA
survey and $4.0\times10^7$ M$_{\sun}$ Mpc$^{-3}$ for the summation
$(\Sigma M_{\rm \HI}/V_{\rm max})$ (H$_{\circ}=75$ km s$^{-1}$ Mpc$^{-1}$)
over the A\HI\/SS galaxies. 

The shallow ZOA and A\HI\/SS control samples are not ideal. The shallow
ZOA survey itself includes the GA region as well as the Local Void and
is smaller than the present survey. The A\HI\/SS sample is smaller still.
The errors in calculating overdensity are therefore dominated by the
control samples.  Until \HI\/PASS solves this particular problem, we will
use a value for the GA overdensity of $\delta
\rho/\rho_{\circ}=0.5\pm0.2$. This is less than the overdensity of
$\sim 1.2$ predicted by POTENT at 4000 km s$^{-1}$ (Kolatt et al.
1995). Around half the \HI~mass lies between $10^9$ and $10^{10}$
 M$_{\sun}$. At this mass, the average maximum distance is $\sim 90$
Mpc, and the weighted sample volume ($n/\Sigma V_{\rm max}^{-1}$) is
$\sim 2\times 10^4$ Mpc$^3$.  Therefore, the excess \HI~mass in this
region is $5.3\times 10^{11}$ M$_{\sun}$.  Assuming \HI~traces total
mass, the excess total mass is $\sim2\times 10^{15} \Omega_{\circ}$
M$_{\sun}$.

Although the \HI~in the GA region appears to trace significant excess
mass, the amount involved is no more than is likely to be present in
the ACO 3627 cluster, which has a mass of $\sim 10^{15}$ M$_{\odot}$ (Woudt
1998; B\"{o}hringer et al. 1996). The combination is well short of the
$5\times 10^{16}$ M$_{\odot}$ invoked to explain local
galaxy dynamics (Lynden-Bell et al. 1988).

\section{Where is the Great Attractor?}

\begin{figure}[htb]
    \plotone{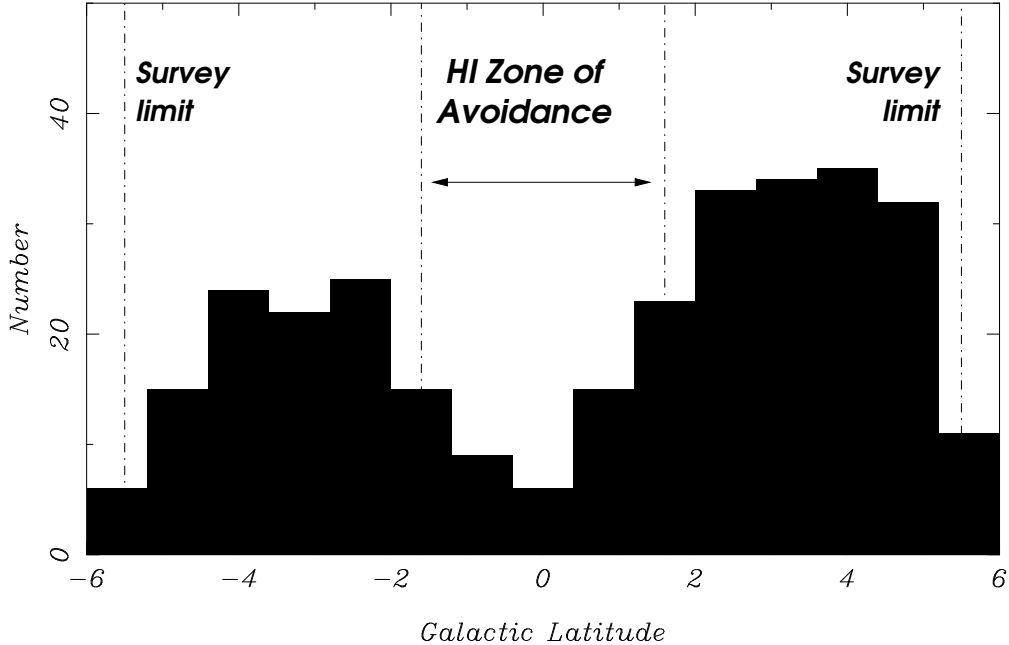}
    \caption{The distribution of galaxies in Galactic latitude shows
an excess of galaxies on the northern side of the Plane relative to
the southern side, and a deficit of galaxies within 1\fdg6 of
the Plane. The deficit is due to increased system temperature and excess
baseline ripple which reduce sensitivity.}
\end{figure}

We now consider alternative hiding places and alternative hypotheses to the
GA:

First, galaxy clusters tend to be deficient in neutral gas, so \HI~
surveys are not efficient cluster detectors. Could the GA consist of
10-50 clusters hidden within this region? Unfortunately, Fig.~1 shows
that the massive superclusters around rich clusters such as ACO 3627
are difficult to hide, and that only a few such superclusters can be
present in this region.

Second, the sensitivity of the current survey is lowered near the
Plane due to increased noise and baseline curvature due to Galactic
continuum sources. How wide is this region and would it be able to
hide significant mass? Figure 4 shows the distribution, in Galactic
latitude, of the 305 galaxies. Although galaxies are present at
$b=0\deg$, there appears to be a real deficit (by a factor of $\sim
2$) at $|b|<1\fdg6$.  This deficit is accounted for in the mass function,
but there is nevertheless some room for hiding a compact cluster. At a
distance of $\sim 70$ Mpc, the remaining hidden zone corresponds to a
projected diameter of $\sim 4$ Mpc. As already mentioned, further
suppression of continuum is planned in order to reduce the
remaining ZOA. At this stage, however, there seems little chance that
important structures have been entirely missed.

Third, could the GA be located outside the present area, or be more
extensive than the region surveyed so far? There is certainly room for
hiding massive structures near the Plane at $\ell<332\deg$ and
$\ell>300\deg$, which have not yet been completely observed by the
Parkes telescope. A GA of around $20\times$ the extent of the
present volume, or lying beyond the redshift of maximum sensitivity
would also help. However, regions away from the Plane are adequately
mapped by IRAS redshift surveys (Saunders et al., these
proceedings) which already fail to completely explain the degree of
overdensity predicted by POTENT.

Finally, there remains the possibility that properties such as bias
parameter and galactic mass-to-light ratios vary on a large-scale in
such a way as to mimic dynamical effects. Comparison of a number of
different distance indicators tends to suggest this is not the case,
and that the issue may be errors and biases. It is notable that the
most accurate distance indicator in the redshift range considered
here, the surface brightness fluctuation method, appears to favour low
values for large-scale bulk flow and a low value for the Great
Attractor mass ($\sim 9\times 10^{15}$ M$_{\odot}$, Tonry et al.
2000), which is much more in accord with the results presented here.

\section{Summary}

An ongoing deep survey with the Parkes telescope has so far found 305
galaxies deep in the Zone of Avoidance at Galactic latitudes
$|b|<5\deg$, longitudes $300\deg < \ell < 332\deg$. These galaxies
fill in the main missing gap in our knowledge of structure in the
local Universe out to $10^4$ km s$^{-1}$ in the `Great Attractor'
region. The excess mass found, $\sim2\times 10^{15} \Omega_{\circ}$
M$_{\sun}$, is significant but much less than that predicted by some
reconstructions of the non-Hubble velocity field (Kolatt et al. 1995).
Several structures (filaments, voids and superclusters) are identified.

\acknowledgements We wish to thank the rest of the ZOA team (A. J. Green,
R. D. Ekers, R. F. Haynes, R. M. Price, E. M. Sadler and
A. Schr\"{o}der) for their contributions.
The research of PH is supported by NSF Faculty Early Career Development
(CAREER) Program award AST 95-02268.  PH also thanks the CSIRO for
support and hospitality during her sabbatical stay.
RCKK thanks CONACyT for their support (research grant 27602E).

\end{document}